\documentclass[aps,preprint,showpacs]{revtex4}
\usepackage{graphicx}
\usepackage{graphics}
\usepackage{epsfig}
\begin{document}
\title{Reply to "Comment on 'Thomson rings in a disk' "
}
\author{A. Puente}
\affiliation{
Departament de F{\'\i}sica,
 Universitat de les Illes Balears, E-07122 Palma de Mallorca, Spain}
\author{R.G. Nazmitdinov}
\affiliation{
Departament de F{\'\i}sica,
 Universitat de les Illes Balears, E-07122 Palma de Mallorca, Spain}
\affiliation{Bogoliubov Laboratory of Theoretical Physics,
Joint Institute for Nuclear Research, 141980 Dubna, Russia}
\author{M. Cerkaski}
\affiliation{Department of Theory of Structure of Matter,
Institute of Nuclear Physics PAN, 31-342 Cracow, Poland}
\author{K.N. Pichugin}
\affiliation{
 Kirensky Institute of Physics, 660036 Krasnoyarsk, Russia}
\begin{abstract}
We demonstrate that our model [Phys.Rev. E{\bf 91}, 032312 (2015] serves
as a useful tool
  to trace the evolution of equilibrium configurations of
one-component charged particles confined in  a disk.
Our approach  reduces significantly the computational effort
in minimizing the energy of equilibrium configurations and
demonstrates a remarkable agreement with the values provided by molecular dynamics
calculations.
We show that the comment
misrepresents our paper, and fails to provide plausible arguments against the
formation hexagonal structure for $n\ge 200$ in molecular dynamics calculations.
\end{abstract}
\date{\today}
\maketitle

In our recent publication \cite{cnp}, we have developed a semi-analytical
approach  that allows to determine equilibrium configurations
for arbitrary, but finite, number of charged particles confined in a disk geometry.
In the Comment \cite{am} by Amore it was found that the minimum energy
configuration of $N=395$ charges
confined in disk and interacting via the Coulomb potential
has  a lower energy than the result
of our molecular dynamics (MD) calculations \cite{cnp}.
Based only on this result Amore concluded that{\it
"...the formation of a hexagonal core and
valence circular rings for the centered configurations, predicted by
the model of Ref.\cite{cnp}, is not supported by numerical
evidence and the configurations obtained with this model cannot be
used as a guide for the numerical calculations, as claimed by the
authors. In light of this findings, the validity of the model of Ref.\cite{cnp}
must be questioned, particular for $N\gtrsim 200$."}

Hereafter, for the sake of convenience we refer to our model as the circular model (CM).
We agree with the author that his possible global MD minimum is better than our estimate for
the particular case $N=395$. However, this is not enough
to conclude that the CM can not help to reduce substantially the computational effort
in MD or simulated annealing (SA) calculations for the following reasons.

\begin{enumerate}
\item
From the Monte Carlo and MD calculations, even for a relatively small number of charged particles,
it follows that the amount of stable configurations grows very rapidly with the number of particles. 
Sometimes, metastable states with lower (or higher) symmetry are found with much higher probability than
the true ground state. This fact was confirmed by the author
who {\it"generated 3001 configurations ..."} to get just one instance of the improved $E_{\rm MIN}= 110664.44$
new tentative ground state, with our prediction for the particle number at the boundary ring:
{\it"... Np = 147 charges are disposed on the border of the disk, in agreement with Ref.1"}.
Evidently, in contrast to his claim, Amore has confirmed the usefulness of the CM.

Indeed, the particle number on the boundary ring $N_p$ is one of the key elements for any
calculation, since once it is defined, it is necessary to simulate less various configurations
(with a number of charges $N-N_p$).
We recall that $N_p>>N_{p-1}>N_{p-2}>\cdots>1$, where $p$ is a number of rings, and $N$ is
a total number of charges.

In fact, external ring occupations are extremely well predicted with some occasional $\pm 1$ mismatch
by means of the expression $N_p(N) = [ 2.795 N^{2/3} - 3.184 ]$,
where $p\simeq [\sqrt{N}/2]$ \cite{cnp}. It is noteworthy that these expressions are
obtained from the systematic CM results.

\item
In our publication \cite{cnp}, in order to obtain 
our estimate of the MD ground state $E_{\rm MD}$,
we generated only 100 configurations  
with the boundary ring $N_{p=9}=147$ charges,
where the internal charges were randomly distributed.
As a result, we have obtained 
$ E_{\rm CM} = 110667.6>E_{\rm MD} = 110665.1>E_{\rm MIN}= 110664.44$.
Note, however, that the disagreement between the author's new result and 
our model prediction $E_{\rm CM}$ is
still less than $3\times 10^{-3}\%$ (as we stated in our paper it is $2\times 10^{-3}\%$).
Moreover, the occupations for the external (approximately circular) shells are
quite accurately predicted within CM for any $N$. In the case of $N=395$ we 
have obtained $(147,65,50,40)$,
while the analysis of the Amore's MD ground state yields $(147,66,51,40)$.
This comparison  suggests that the effectivity of the CM prediction might be improved
if the second ring, the nearest neigbor to the boundary one, should be taken into account.

\item
To prove the usefulness of this idea we consider 
initial configurations characterized by external occupations: 
$N_9=147$ (Set 1); $N_9=147$, $N_8=65$ (Set 2); 
$N_9=147$, $N_8=66$ (Set 3). 
In all cases we have generated 2000
configurations, where $N_9$ particles were initially set on the
boundary at $R_1=1$, and for two other sets $N_8$ particles 
have been distributed at $R_2=0.96$. That value was chosen to
take into account monopole oscillations around the equilibrium 
configuration. The remaining particles were
distributed randomly. 

For the Set 1 (Fig.\ref{fig1}, top panel) we found 
the lowest state $E_{\rm MD}=110664.52>E_{\rm MIN}= 110664.44$, that occurs just once. 
In the middle panel (Set 2) we use two boundary shells $N_9=147$, $N_8=65$, 
predicted by the CM partition,
and obtain slightly lower state. However, the ground state is not reached yet. 

The systematic analysis of the CM results 
leads us to conclusion that  the second shell occupation is fitted by 
the formula $N_{p-1}(N)= [1.351 N^{2/3} - 6.566 ]$ that yields $N_8=66$. 
Considering the initial configuration
with $N_9=147$, $N_8=66$ (Set 3) with randomly distributed internal charges
(Fig.\ref{fig1}, bottom panel), we 
obtain that the ground state $E_{\rm MIN}$ occurs three times
$(0.15\%$).
In other words, with this initialization it appears once every 666 generated 
configurations. Note that Amore has generated 3001 configurations to obtain
 just one realization of the possible ground state.
 \begin{center}
\begin{figure}
\hspace{3cm}
\includegraphics[scale=0.75]{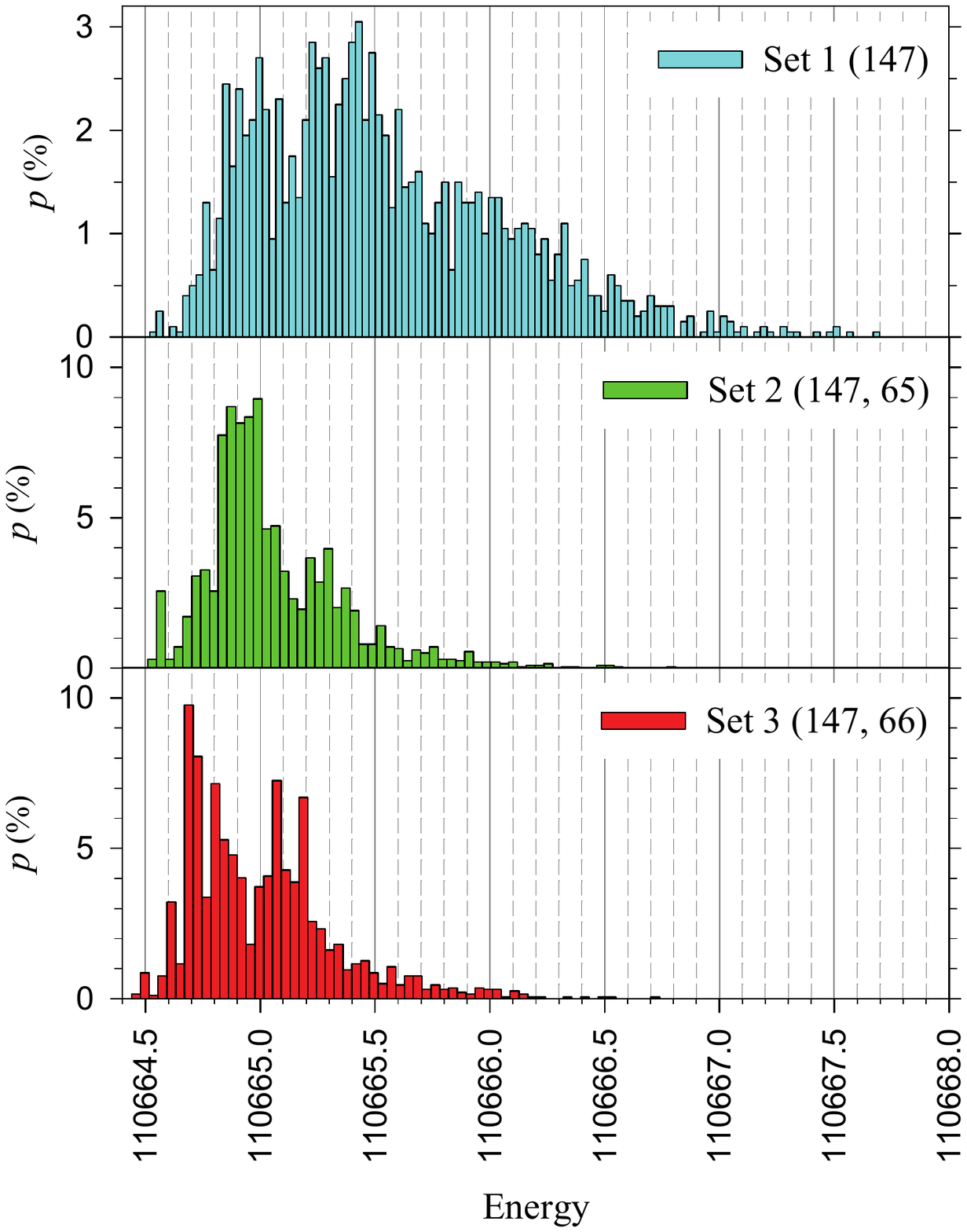}
\vspace{-7.cm}
  \caption{(Color online) 
  Histograms for energy states of $N=395$
  charges in the disk geometry obtained by means of the MD method
  for different initialization procedures.}
    \label{fig1}
\end{figure}
\end{center}

\item
We recall that for infinite systems the hexagonal lattice has the lowest energy of
all two-dimensional Wigner Bravais crystals \cite{mar}. Evidently,
the decrease of system size places primary emphasis upon system boundaries
(see, for example, discussion in Refs.\cite{fin,bir}).
Therefore, one needs to understand how the Wigner
crystallization may settle down, in a particular, in a disk geometry
as a function of the number of interacting
charged particles. In Supplemental Material we compare
our results corresponding to the MD
and the semi-analytical approach for $161\leq N\leq 260$ charges.
These results demonstrate a remarkable agreement between two approaches and
make it clear how the centered hexagonal lattice (CHL)
evolves with the increase of charge particle number.
Therefore, we strongly believe that the results obtained by means of our method can be successfully used to
feed SA or MD calculations with sensible initial configurations, reducing significantly the amount of
scanning, normally needed to visit the global energy minima.

\item
The systematic manifestation of the CHL
with the increase of particle number $N\geq 200$ in our CM and MD results
can be interpreted as the onset of the hexagonal crystallization in the disk geometry.
There is no, however, any manifestation of a phase transition, typical for infinite systems.
In  a finite system a crossover takes place from the CHL
to ring localization with the approaching to the disk boundary.
This ring organization is clearly seen  at the boundary in Fig.2 presented by the author (two clear rings).
\end{enumerate}

In our paper we have compared the MD configuration with the prediction of the CM for the CHL
at $N=395$.
In fact, in our MD calculations the clean CHL  takes place at $N=381$ with the configuration
${\bf 143,64,49},39,30,19,{\bf 18,12,6,1}$ and the minimal energy $E=102764.53$.
The valence configurations with $N_7=49$, $N_8=64, N_9=143$ form well defined ring structure.

The increase of particle number disintegrates slowly the CHL in the disk geometry,
while the hexagonal lattice still exists. Nevertheless, with each new shell, as soon as 
a new particle appears at the centre it gives rise to the CHL again. Since we deal with 
a finite system, restricted by the circular geometry, the boundary affects the plain 
symmetrical configurations giving rise to defects.

In conclusion, we disagree with the main outcome of the author's Comment formulated in his Summary.
In order to  argue against our model and corresponding conclusions, it is required a systematic
thorough analysis of the system with increasing particle number but not only one particular case.
In fact, we have demonstrated that the CM predictions for external rings $(N_p, N_{p-1})$ enable to us to 
reduce substantially  scanning efforts needed to reach the ground state in the MD. 

\section*{Acknowledgments}
M.C. and K.P. are grateful for the warm
hospitality at JINR. This work was supported in
part by Bogoliubov-Infeld program of BLTP and
Russian Foundation for Basic Research.

\newpage
\centerline{\bf Supplemental Material}
\begin{center}
{\bf \large The MD and Circular Model results}
\end{center}

 These tables summarize our results corresponding to
 the minimum energy equilibrium configurations under disk confinement. $E_{\rm MD}$,
 $E_{\rm CM}$ are the total MD (our best estimate) and the Circular Model energies. 
\begin{table}[h]
\centering
{\LARGE  Results for $161\le n \le 180$\par}
\begin{tabular} {@{\hspace{0.15cm}} c @{\hspace{0.5cm}} c @{\hspace{0.5cm}} c @{\hspace{0.25cm}} | @{\hspace{0.25cm}} c @{\hspace{0.5cm}} c @{\hspace{0.5cm}}}
\hline
$n$  & $E_{\rm MD}$ &  Configuration & $E_{\rm CM}$ & Configuration\\
\hline
161 & 17323.885 & [79,33,22,15,9,3] & 17324.571 & [79,33,22,15,9,3]\\
162 & 17548.672 & [79,33,23,15,9,3] & 17549.325 & [79,33,23,15,9,3]\\
163 & 17775.066 & [80,33,23,15,9,3] & 17775.717 & [80,33,23,15,9,3]\\
164 & 18002.880 & [80,33,23,16,9,3] & 18003.488 & [80,33,23,16,9,3]\\
165 & 18232.249 & [80,34,23,16,9,3] & 18232.837 & [80,34,23,16,9,3]\\
166 & 18463.081 & [81,34,23,16,9,3] & 18463.668 & [81,34,23,16,9,3]\\
167 & 18695.335 & [81,34,24,16,9,3] & 18695.901 & [81,34,24,16,9,3]\\
168 & 18929.198 & [81,34,24,16,10,3] & 18929.751 & [81,34,24,16,10,3]\\
169 & 19164.419 & [81,34,24,16,10,4] & 19164.969 & [81,34,24,16,10,4]\\
170 & 19400.980 & [82,34,24,16,10,4] & 19401.528 & [82,34,24,16,10,4]\\
171 & 19639.241 & [82,35,24,16,10,4] & 19639.768 & [82,35,24,16,10,4]\\
172 & 19878.944 & [83,35,24,16,10,4] & 19879.461 & [83,34,24,17,10,4]\\
173 & 20120.029 & [83,35,24,17,10,4] & 20120.507 & [83,35,24,17,10,4]\\
174 & 20362.731 & [83,35,25,17,10,4] & 20363.180 & [83,35,25,17,10,4]\\
175 & 20606.871 & [84,35,25,17,10,4] & 20607.318 & [84,35,25,17,10,4]\\
176 & 20852.540 & [84,35,25,17,11,4] & 20852.999 & [84,36,25,17,10,4]\\
177 & 21099.703 & [84,36,25,17,11,4] & 21100.164 & [84,36,25,17,11,4]\\
178 & 21348.301 & [85,36,25,17,11,4] & 21348.762 & [85,36,25,17,11,4]\\
179 & 21598.311 & [85,36,25,18,11,4] & 21598.791 & [85,36,25,17,11,5]\\
180 & 21849.924 & [85,36,25,18,11,5] & 21850.334 & [85,36,25,18,11,5]\\
\hline
\end{tabular}
\end{table}
\vskip 5mm

\begin{table}[h]
\centering
{\LARGE  Results for $181\le n \le 207$\par}
\begin{tabular} {@{\hspace{0.15cm}} c @{\hspace{0.5cm}} c @{\hspace{0.5cm}} c @{\hspace{0.25cm}} | @{\hspace{0.25cm}} c @{\hspace{0.5cm}} c @{\hspace{0.5cm}}}
\hline
$n$  & $E_{\rm MD}$ &  Configuration & $E_{\rm CM}$ & Configuration\\
\hline
181 & 22102.961 & [86,36,25,18,11,5] & 22103.368 & [86,36,25,18,11,5]\\
182 & 22357.440 & [86,36,26,18,11,5] & 22357.815 & [86,36,26,18,11,5]\\
183 & 22613.517 & [86,37,26,18,11,5] & 22613.878 & [86,37,26,18,11,5]\\
184 & 22871.031 & [87,37,26,18,11,5] & 22871.391 & [87,37,26,18,11,5]\\
185 & 23129.918 & [87,36,26,18,11,6,1] & 23130.442 & [87,37,26,18,12,5]\\
186 & 23390.285 & [87,37,26,18,11,6,1] & 23391.044 & [87,37,26,19,12,5]\\
187 & 23652.188 & [87,37,26,18,12,6,1] & 23652.947 & [87,37,26,18,12,6,1]\\
188 & 23915.459 & [88,37,26,18,12,6,1] & 23916.215 & [88,37,26,18,12,6,1]\\
189 & 24180.381 & [88,37,27,18,12,6,1] & 24181.062 & [88,37,26,19,12,6,1]\\
190 & 24446.798 & [89,37,27,18,12,6,1] & 24447.458 & [88,37,27,19,12,6,1]\\
191 & 24714.561 & [89,37,27,19,12,6,1] & 24715.189 & [89,37,27,19,12,6,1]\\
192 & 24983.853 & [89,38,27,19,12,6,1] & 24984.461 & [89,38,27,19,12,6,1]\\
193 & 25254.755 & [90,38,27,19,12,6,1] & 25255.362 & [90,38,27,19,12,6,1]\\
194 & 25527.119 & [90,38,27,19,13,6,1] & 25527.697 & [90,38,27,19,13,6,1]\\
195 & 25801.014 & [90,38,27,20,13,6,1] & 25801.595 & [90,38,28,19,13,6,1]\\
196 & 26076.378 & [91,38,27,20,13,6,1] & 26076.964 & [91,38,28,19,13,6,1]\\
197 & 26353.122 & [91,39,27,20,13,6,1] & 26353.672 & [91,38,27,20,13,7,1]\\
198 & 26631.365 & [91,39,28,20,13,6,1] & 26631.870 & [91,39,27,20,13,7,1]\\
199 & 26911.103 & [91,39,28,20,13,7,1] & 26911.559 & [91,39,28,20,13,7,1]\\
200 & 27192.287 & [92,39,28,20,13,7,1] & 27192.741 & [92,39,28,20,13,7,1]\\
201 & 27475.149 & [92,40,28,20,13,7,1] & 27475.591 & [92,40,28,20,13,7,1]\\
202 & 27759.495 & [92,39,28,21,14,7,1] & 27759.953 & [93,40,28,20,13,7,1]\\
203 & 28045.151 & [93,39,28,21,14,7,1] & 28045.669 & [93,40,29,20,13,7,1]\\
204 & 28332.320 & [93,40,28,21,14,7,1] & 28332.900 & [93,40,29,20,14,7,1]\\
205 & 28621.069 & [93,40,29,21,14,7,1] & 28621.647 & [93,40,29,21,14,7,1]\\
206 & 28911.236 & [94,40,29,21,14,7,1] & 28911.813 & [94,40,29,21,14,7,1]\\
207 & 29203.054 & [94,41,29,21,14,7,1] & 29203.620 & [94,41,29,21,14,7,1]\\
\hline
\end{tabular}
\end{table}
\vskip 5mm

\begin{table}
\centering
{\LARGE  Results for $208\le n \le 234$\par}
\begin{tabular} {@{\hspace{0.15cm}} c @{\hspace{0.5cm}} c @{\hspace{0.5cm}} c @{\hspace{0.25cm}} | @{\hspace{0.25cm}} c @{\hspace{0.5cm}} c @{\hspace{0.5cm}}}
\hline
$n$  & $E_{\rm MD}$ &  Configuration & $E_{\rm CM}$ & Configuration\\
\hline
208 & 29496.341 & [94,40,29,21,14,8,2] & 29496.944 & [94,41,29,21,14,8,1]\\
209 & 29790.985 & [95,40,29,21,14,8,2] & 29791.618 & [95,41,29,21,14,8,1]\\
210 & 30087.107 & [95,41,29,21,14,8,2] & 30087.834 & [95,41,30,21,14,8,1]\\
211 & 30384.840 & [95,41,30,21,14,8,2] & 30385.659 & [95,41,30,22,14,8,1]\\
212 & 30684.005 & [96,41,30,21,14,8,2] & 30684.828 & [96,41,30,22,14,8,1]\\
213 & 30984.546 & [96,41,30,22,14,8,2] & 30985.505 & [96,41,30,22,15,8,1]\\
214 & 31286.636 & [96,41,30,22,13,9,3] & 31287.674 & [96,41,30,22,15,8,2]\\
215 & 31590.285 & [97,41,30,22,13,9,3] & 31591.327 & [97,41,30,22,15,8,2]\\
216 & 31895.274 & [97,41,30,22,14,9,3] & 31896.396 & [97,42,30,22,15,8,2]\\
217 & 32201.810 & [97,41,30,22,15,9,3] & 32202.828 & [97,41,30,22,15,9,3]\\
218 & 32509.860 & [97,42,30,22,15,9,3] & 32510.843 & [97,42,30,22,15,9,3]\\
219 & 32819.357 & [98,42,30,22,15,9,3] & 32820.338 & [98,42,30,22,15,9,3]\\
220 & 33130.391 & [98,42,31,22,15,9,3] & 33131.367 & [98,42,31,22,15,9,3]\\
221 & 33443.063 & [98,42,31,23,15,9,3] & 33444.031 & [98,42,31,23,15,9,3]\\
222 & 33757.067 & [99,42,31,23,15,9,3] & 33758.034 & [99,42,31,23,15,9,3]\\
223 & 34072.594 & [99,42,31,23,16,9,3] & 34073.531 & [99,42,31,23,16,9,3]\\
224 & 34389.617 & [99,43,31,23,16,9,3] & 34390.523 & [99,43,31,23,16,9,3]\\
225 & 34708.151 & [100,43,31,23,16,9,3] & 34709.055 & [100,43,31,23,16,9,3]\\
226 & 35028.160 & [100,43,31,23,15,10,4] & 35029.106 & [100,43,31,23,16,9,4]\\
227 & 35349.639 & [100,43,31,23,16,10,4] & 35350.486 & [100,43,31,23,16,10,4]\\
228 & 35672.675 & [101,43,31,23,16,10,4] & 35673.518 & [101,43,31,23,16,10,4]\\
229 & 35997.082 & [101,43,32,23,16,10,4] & 35997.893 & [101,43,32,23,16,10,4]\\
230 & 36323.076 & [101,44,32,23,16,10,4] & 36323.861 & [101,44,32,23,16,10,4]\\
231 & 36650.589 & [101,44,32,24,16,10,4] & 36651.395 & [101,44,32,24,16,10,4]\\
232 & 36979.503 & [102,44,32,24,16,10,4] & 36980.306 & [102,44,32,24,16,10,4]\\
233 & 37309.955 & [102,44,32,24,17,10,4] & 37310.715 & [102,44,32,24,17,10,4]\\
234 & 37642.049 & [102,44,33,24,17,10,4] & 37642.792 & [102,44,33,24,17,10,4]\\
\hline
\end{tabular}
\end{table}
\vskip 5mm

\begin{table}
\centering
{\LARGE  Results for $235\le n \le 260$\par}
\begin{tabular} {@{\hspace{0.15cm}} c @{\hspace{0.5cm}} c @{\hspace{0.5cm}} c @{\hspace{0.25cm}} | @{\hspace{0.25cm}} c @{\hspace{0.5cm}} c @{\hspace{0.5cm}}}
\hline
$n$  & $E_{\rm MD}$ &  Configuration & $E_{\rm CM}$ & Configuration\\
\hline
235 & 37975.484 & [103,44,33,24,17,10,4] & 37976.225 & [103,44,33,24,17,10,4]\\
236 & 38310.462 & [103,44,33,24,17,11,4] & 38311.186 & [103,45,33,24,17,10,4]\\
237 & 38646.933 & [103,45,33,24,17,11,4] & 38647.705 & [103,45,33,24,17,11,4]\\
238 & 38984.913 & [104,45,33,24,17,11,4] & 38985.683 & [104,45,33,24,17,11,4]\\
239 & 39324.318 & [104,45,33,25,16,11,5] & 39325.033 & [104,45,33,24,17,11,5]\\
240 & 39665.177 & [104,45,33,25,14,12,6,1] & 39665.976 & [104,45,33,25,17,11,5]\\
241 & 40007.632 & [104,45,33,25,15,12,6,1] & 40008.473 & [105,45,33,25,17,11,5]\\
242 & 40351.460 & [105,45,33,25,15,12,6,1] & 40352.337 & [105,45,33,25,18,11,5]\\
243 & 40696.877 & [105,45,33,25,16,12,6,1] & 40697.757 & [105,46,33,25,18,11,5]\\
244 & 41043.791 & [105,46,33,25,16,12,6,1] & 41044.696 & [105,46,34,25,18,11,5]\\
245 & 41392.174 & [106,46,33,25,16,12,6,1] & 41393.091 & [106,46,34,25,18,11,5]\\
246 & 41741.996 & [106,46,34,25,16,12,6,1] & 41743.132 & [106,46,34,25,18,12,5]\\
247 & 42093.362 & [106,46,34,25,17,12,6,1] & 42094.661 & [106,46,34,25,18,11,6,1]\\
248 & 42446.278 & [107,46,34,25,17,12,6,1] & 42447.440 & [106,46,34,25,18,12,6,1]\\
249 & 42800.557 & [107,46,34,25,18,12,6,1] & 42801.689 & [107,46,34,25,18,12,6,1]\\
250 & 43156.448 & [107,46,34,26,18,12,6,1] & 43157.543 & [107,46,34,26,18,12,6,1]\\
251 & 43513.864 & [107,47,34,26,18,12,6,1] & 43514.922 & [107,47,34,26,18,12,6,1]\\
252 & 43872.683 & [108,47,34,26,18,12,6,1] & 43873.737 & [108,47,34,26,18,12,6,1]\\
253 & 44233.025 & [108,47,35,26,18,12,6,1] & 44234.016 & [108,47,34,26,19,12,6,1]\\
254 & 44594.889 & [108,47,35,26,19,12,6,1] & 44595.833 & [108,47,35,26,19,12,6,1]\\
255 & 44958.250 & [109,47,35,26,19,12,6,1] & 44959.191 & [109,47,35,26,19,12,6,1]\\
256 & 45323.172 & [109,47,35,26,19,13,6,1] & 45324.102 & [109,48,35,26,19,12,6,1]\\
257 & 45689.578 & [109,48,35,26,19,13,6,1] & 45690.514 & [109,48,35,26,19,13,6,1]\\
258 & 46057.509 & [110,48,35,26,19,13,6,1] & 46058.440 & [110,48,35,26,19,13,6,1]\\
259 & 46426.854 & [110,48,35,27,19,13,6,1] & 46427.711 & [110,48,35,26,19,13,7,1]\\
260 & 46797.668 & [110,48,35,27,19,13,7,1] & 46798.489 & [110,48,35,27,19,13,7,1]\\
\hline
\end{tabular}
\end{table}

\end{document}